# 64-GBd DP-Bipolar-8ASK Transmission over 120 km SSMF Employing a Monolithically Integrated Driver and MZM in 0.25-µm SiGe BiCMOS Technology

Gilda Raoof Mehrpoor[1,4], Carsten Schmidt-Langhorst[2], Benjamin Wohlfeil[1], Robert Elschner[2], Danish Rafique[1], Robert Emmerich[2], Annika Dochhan[1], Iria Lopez[3], Pedro Rito[3], Despoina Petousi[3], Dietmar Kissinger[3], Lars Zimmermann[3], Colja Schubert[2], Bernhard Schmauss[4], Michael Eiselt[1], Jörg-Peter Elbers[1]

[1]*ADVA Optical Networking SE, Fraunhoferstr. 9a, 82152 Martinsried/Munich, Germany*
[2]*Fraunhofer Heinrich Hertz Institute, Einsteinufer 37, 10587 Berlin, Germany*
[3]*IHP GmbH, Im Technologiepark 25, 15236 Frankfurt (Oder), Germany*
[4]*LHFT, Friedrich-Alexander Universität Erlangen-Nürnberg, Cauerstr. 9, 91058 Erlangen, Germany*
*Author e-mail address: gmehrpoor@advaoptical.com*

**Abstract:** We demonstrate 64-GBd signal generation up to bipolar-8-ASK utilizing a single MZM, monolithically integrated with segmented drivers in SiGe. Using polarization multiplexing, 300-Gb/s net data rate transmission over 120 km SSMF is shown.
**OCIS codes:** (060.1660) Coherent communication; (130.0250) Optoelectronics; (130.4110) Modulators

## 1. Introduction

Upcoming coherent transceivers will support 64-GBd polarization-division multiplexed 16-ary quadrature-amplitude modulation (DP-16QAM) to achieve a line rate of 400 Gb/s per optical carrier [1]. Increasing the cardinality to DP-64QAM is a promising pathway towards a line rate of 600 Gb/s per carrier.

Electronic-photonic integrated circuits (EPIC) comprising passive electro-optical components and high-speed radio-frequency (RF) electronics monolithically integrated on silicon enable a high level of integration and thus potentially offer compact, low cost and power efficient transceiver solutions [2]-[4]. Recently, a depletion-type silicon Mach-Zehnder modulator (Si-MZM) with segmented linear driver amplifiers, monolithically integrated in a photonic BiCMOS technology EPIC, was demonstrated with modulation bandwidths on the order of 40 GHz. RF drivers of high linearity enabled the generation and transmission of 112-Gb/s PAM4 signals using direct detection [5].

In this paper, we present for the first time 64-GBd coherent system experiments with this type of modulator EPIC. As a first step towards QAM modulation, we generate 64-GBd bipolar $m$-ary amplitude-shift keying (Bi-$m$ASK) signals with cardinalities of $m = 2$, 4, and 8. Using polarization multiplexing, this yields DP-Bi-$m$ASK signals with gross line rates of 128 Gb/s, 256 Gb/s and 384 Gb/s, respectively. Employing the DP-Bi-8ASK format and coherent detection, we accomplish transmission over 120 km of standard single-mode fiber (SSMF) at a net data rate of 300 Gb/s with >2.5 dB-Q margin to the threshold for soft-decision forward-error correction (SD-FEC).

## 2. Monolithically integrated segmented driver and MZM in 0.25-µm SiGe BiCMOS

Due to weak electro-optical effects at telecom wavelengths in silicon, signal modulation is typically attained by means of charge density variation in material. Applying a reverse bias voltage to a rib waveguide based diode structure results in a phase modulation of the light signal propagating within the depletion area [4]. The variation of real and imaginary parts of the refractive index, induced by the applied electric field, causes an additional, unwanted intensity modulation, the impact of which needs to be assessed.

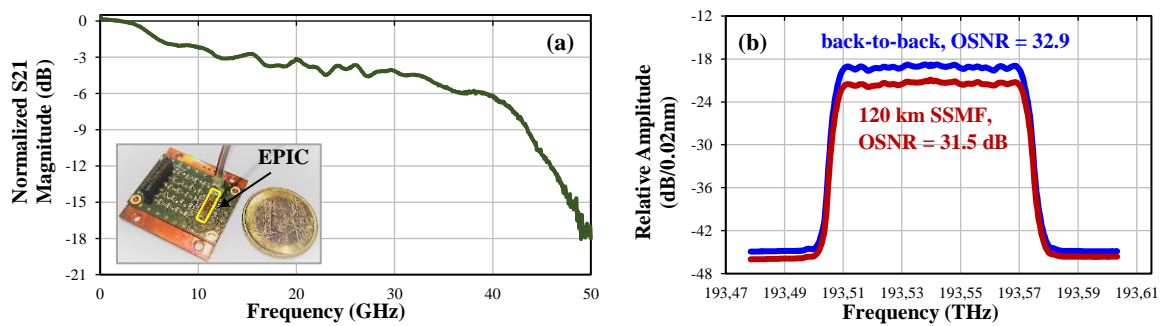

**Fig. 1**: (a) Measured frequency response of the device under test; Inset: Photograph of Si-chip in PCB;
(b) Optical spectra for 64-GBd DP-bipolar-8ASK in b2b case and after 120 km SSMF transmission.

The MZM structure used in this work is based on two parallel 6 mm long phase shifters requiring a $V_\pi$ of 4.5 V. The input laser light is split, and combined at the output, via 2×2 Multi-Mode-Interference couplers (MMI). In order to compensate the initial phase imbalances between the two arms, as well as to define the operating point of the MZM, 1 mm Ohmic heaters are implemented on each arm and tuned using an external DC voltage [4]. The EPIC is fabricated in 0.25 µm SiGe BiCMOS technology, comprising 16 phase shifter segments, connected to a segmented dual-stage driver with adjustable gain (up to 14.5 dB), and leading to optimal delay matching, reduced RF loss and high voltage swing. The reverse bias of the MZM waveguide diodes is set by an external DC voltage on the p-side of both arms.

For test purposes, the fabricated EPIC (9.8 mm×1.3 mm area) was mounted on a PCB as shown in **Fig. 1a** (inset), where the light was coupled through a fiber array into the input and output 1D grating couplers, while the high-speed differential RF signal is coupled via an RF probe. The total static insertion loss of the device is 18 dB, of which 8 dB is due to the grating couplers. The measured S21 frequency response of the EPIC device (**Fig. 1a**) yields 3-dB and 6-dB bandwidths of 11 GHz and 35 GHz, respectively. The S21 data was used for a linear transmit-side pre-distortion of the RF drive signal. Exemplary optical spectra of a 64-GBd DP-Bi-8ASK signal generated through the EPIC are illustrated in **Fig. 1b** for the back-to-back (b2b) configuration at the maximum available optical signal-to-noise ratio (OSNR, measured in 0.1 nm) as well as after 120-km SSMF transmission.

## 3. Experimental setup

The experimental set-up is schematically depicted in **Fig. 2**. The 64-GBd dual-polarization data signal is generated from an external cavity laser (ECL, 193.4 THz, +16 dBm) using the monolithically integrated Si-driver-MZM EPIC and a subsequent polarization multiplexing emulation stage (PolMUX, 17.1 ns decorrelation corresponding to 1094 64-GBd symbols). The Si-EPIC is differentially driven utilizing two channels of an 84-GS/s digital-to-analog converter (DAC). The DAC waveforms were pre-processed utilizing offline transmitter-side digital signal processing (TX-DSP) including Bi-2-/4-/8-ary ASK symbol generation (34676 symbols) from random bit sequences, header insertion (incl. binary PRBS-5 training sequences mapped to the outer amplitude levels), pulse shaping (Nyquist root-raised cosine 0.1), up-sampling and linear pre-distortion of the frequency responses of the DAC and the Si-driver-MZM chip. For bipolar signal modulation at the null operation point of the MZM, the heater voltage on one MZM arm was set to 2.1 V. A reverse bias of the RF phase shifter diodes of 2 V was set while the driver gain and operating point was adjusted by its bias voltage of about 3.4 V.

A variable optical attenuator (VOA) was used between transmitter and receiver to either adjust the fiber launch power $P_{launch}$ into the 120-km SSMF link or the OSNR in the b2b case. At the coherent receiver the signal is pre-amplified and filtered (1.4 nm) before being mixed with a local oscillator (LO) in a polarization diversity optical 90° hybrid and detected via four balanced photodetectors (BPD). The outputs are digitized by an 80 GS/s, 33-GHz 4-Ch real-time oscilloscope. The bit-error ratio (BER) was obtained by processing about 1 million acquired samples using standard offline DSP [6], including optional digital compensation of the fiber dispersion, data-aided 2x2 MIMO equalization and blind-phase search carrier phase estimation, followed by a T-spaced 4x4 decision-directed equalizer. Note that the carrier phase was fully recovered and the symbol decision was based solely on the in-phase component of each symbol. Finally, the measured BER was converted to Q²-factors by $Q^2 = 20 \cdot \log_{10}\left[\sqrt{2} \cdot \text{erfc}^{-1}(2 \cdot BER)\right]$.

## 4. Results and discussion

The optical spectrum at the output of the 64-GBd DP-Bi-*m*ASK transmitter is shown **Fig. 1b,** exemplarily for DP-Bi-8ASK modulation at the maximum transmitter OSNR of 32.9 dB (blue). Owing to the transmitter predistortion, the spectrum exhibits only a small ripple (<1 dB). **Figs. 3a-c** show the received constellations after offline DSP at the maximum available OSNRs of 34.0 dB, 32.5 dB and 32.9 dB, respectively. The difference in OSNR is attributed to different modulation-induced MZM losses. The horizontal histogram plots shown together with the constellations show an excellent e/o linearity of the Si-EPIC, i.e. each symbol is modulated at its target in-phase (real) amplitude value. Moreover, each symbol distribution is in very good agreement with a Gaussian fit.

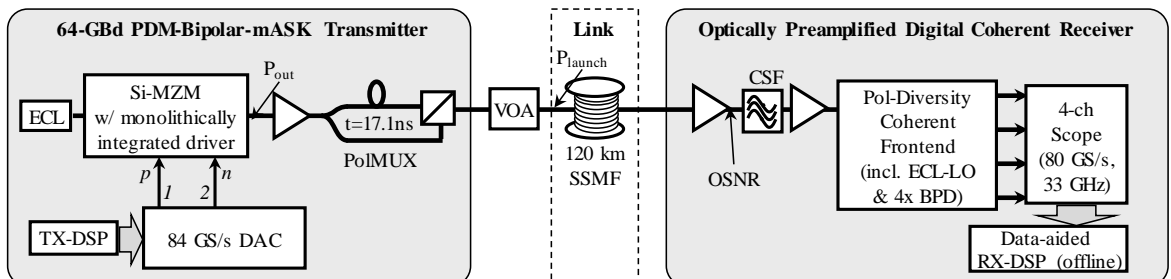

**Fig. 2:** Experimental set-up. CSF: Channel selection filter

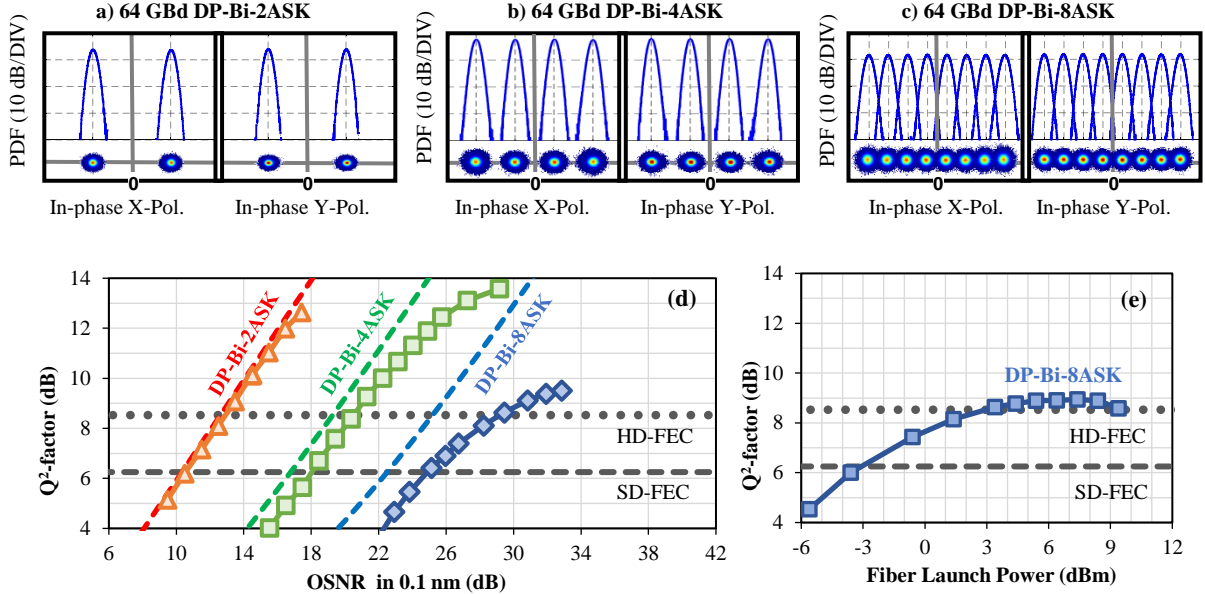

**Fig. 3:** (a-c) back-to-back constellation for DP-Bi-2/4/8-ASK together with histograms along in-phase axis; (d) $Q^2$-factor performance for 64 GBd DP-Bi-2/4/8-ASK (dashed lines: AWGN theory); e) Launch power sweep for 64-GBd DP-Bi-8ASK transmission over 120 km SSMF.

The measured b2b $Q^2$-factor performance versus OSNR is shown in **Fig. 3d**. OSNR-penalties at the SD-FEC threshold (6.25 dB-Q) are found to be 0.2 dB, 1.4 dB and 3.2 dB for DP-Bi-2ASK, DP-Bi-4ASK and DP-Bi-8ASK, respectively. For DP-Bi-2ASK no error floor was observed up to $Q^2$ of 13 dB. The DP-Bi-4ASK format exhibits an error floor around 14 dB-Q while the floor decreases to about 10 dB-Q for DP-Bi-8ASK. This is attributed to the combined effects of higher constellation cardinality and effective amplitude resolution (ENOB) of the DAC and ADC.

Finally, the 120-km transmission results for 384 Gb/s DP-Bi-8ASK are shown in **Fig. 3e** (see received spectrum in **Fig. 1b** (red)). The optimum launch power is found to be about +7.4 dBm, wherefore a $Q^2$-factor of 8.9 dB is achieved after transmission, bringing about 0.4 dB and 2.7 dB Q-margins to the hard-decision FEC (HD-FEC, 8.53 dB-Q) and SD-FEC (6.25 dB-Q) thresholds, respectively. Considering the Gaussian distributions of the received symbols (**Fig. 3(c)**), an error-free transmitted net data rate of 300 Gb/s can be inferred assuming SD-FEC at a total overhead incl. protocols of 28%. Moreover, the received $Q^2$-factor after 120-km is above the HD-FEC threshold for launch powers ranging from about +4 dBm to +9 dBm, indicating an error-free transmitted net data rate of 342 Gb/s assuming HD-FEC at a total overhead incl. protocols of 12%.

## 5. Conclusion

Exploiting a silicon Mach-Zehnder modulator EPIC with monolithically integrated linear driver amplifiers, 64-GBd Bi-*m*ASK signals with cardinalities of *m* = 2, 4, and 8 were generated. Using a PolMux, corresponding DP-Bi-*m*ASK signals with line rates of 128 Gb/s, 256 Gb/s and 384 Gb/s were obtained with implementation penalties of 0.2 dB, 1.4 dB and 3.2 dB, respectively. Using the DP-Bi-8ASK format we showed transmission over 120 km SSMF with >2.5 dB-Q margin to the SD-FEC threshold, indicating an error-free net rate of 300 Gb/s. These achievements support the feasibility of using such modulators in a nested IQ structure, leading to a DP-64QAM scenario.

## 6. Acknowledgements

This work has been partially funded by German Bundesministerium für Bildung und Forschung (BMBF) under the Celtic projects SENDATE Secure-DCI (16KIS0479, C2015/3-4) and SPEED (13N13748, 13N13744). The authors cordially thank the Fraunhofer IZM, especially Gunnar Boettger and Wojciech Lewoczko-Adamczyk for assembly and mounting of the EPIC on the testboard.